# An open-cell environmental transmission electron microscopy technique for in-situ characterization of samples in aqueous liquid solutions

Barnaby D.A. Levin[1], Diane Haiber[1], Qianlang Liu[1], and Peter A. Crozier[1*]

[1.] School for Engineering of Matter, Transport, and Energy, Arizona State University, Arizona, USA.

*Corresponding Author. Email: crozier@asu.edu

**Abstract**

The desire to image specimens in liquids has led to the development of open cell and closed cell techniques in transmission electron microscopy (TEM). The closed cell approach is currently more common in TEM and has yielded new insights into a number of biological and materials processes in liquid environments. The open cell approach, which requires an environmental TEM, is technically challenging but may be advantageous in certain circumstances due to fewer restrictions on specimen and detector geometry. Here, we demonstrate a novel approach to open cell liquid TEM, in which we use salt particles to facilitate the *in situ* formation of droplets of aqueous solution that envelope specimen particles coloaded with the salt. This is achieved by controlling sample temperature to between 1-10ºC and introducing water vapor to the environmental TEM chamber above the critical pressure for the formation of liquid water on the salt particles. Our use of in-situ hydration enables specimens to be loaded into microscope in a dry state using standard 3 mm TEM grids, and specimens can therefore be prepared using trivial sample preparation techniques. Our future aim will be to combine this technique with an in-situ light source to study photocorrosion in aqueous environments.

**Introduction**

A wide variety of important physical and chemical processes occur in liquid environments. This includes the activities of biological cells, chemical reactions in batteries and fuel cells, corrosion, nanoparticle synthesis, and catalytic reactions such as photocatalytic water-splitting. The desire to image specimens and processes in liquids has led to the development of two principal electron microscopy techniques, as discussed in detail below (de Jonge & Ross, 2011; Ross, 2015).

A closed-cell approach is more commonly used for transmission electron microscopy (TEM). This involves enclosing a specimen in liquid between two electron transparent windows, shielding the liquid from the high-vacuum environment of the TEM. Enclosed cells were first applied to study wet specimens in TEM soon after the development of the TEM itself, (Marton, 1935, Abrams & McBain, 1944a, Abrams & McBain, 1944b) but have been more widely adopted in recent years due to advances in microfabrication methods, and the ability to prepare very thin windows from silicon nitride (Williamson et. al. 2003; Grogan & Bau, 2010; Ring & de Jonge, 2010; Holtz et. al. 2013; Dukes et. al. 2014; Jensen et. al. 2014, Unocic et. al. 2014, Zeng et. al. 2014), or even graphene (Yuk et. al. 2012, Chen et. al. 2013, Park et. al. 2015).

An alternative is to use an open-cell approach, which is more commonly employed in environmental scanning electron microscopy (ESEM) (Danilatos, 1991). This technique requires maintaining a gaseous atmosphere inside the microscope sample chamber above the saturation vapor pressure of the gas, which prevents liquid evaporation from the specimen. The saturation vapor pressure is a function of the type of liquid and varies with temperature. Saturation vapor pressure is shown as a function of temperature for water in Figure 1. Maintaining the required pressure is achieved via a differential pumping system, enabling the pressure inside the specimen chamber to be higher than the pressure around the electron gun, which must remain in high vacuum. Some of the early pioneers of TEM first used open cells containing gas to reduce specimen charging (Krause, 1937, Ruska 1942), and control hydrocarbon contamination (Pashley & Presland, 1962). In the 1970s, researchers began to employ differentially pumped open-cell environmental TEM (ETEM) systems to study wet specimens (Matricardi et. al. 1972, Ward & Mitchell, 1972, Parsons, 1974, Parsons et. al. 1974). These open-cell ETEMs could typically achieve maximum gas pressures typically up to ~20 Torr surrounding the specimen. Some designs reported pressures up to 760 Torr surrounding the specimen, although it should be noted that spatial resolution would have been degraded at higher gas pressures (Parsons 1974). Early open cell ETEMs were primarily used to study biological samples, which had to be kept in their natural hydrated state as they were loaded into the microscope (Parsons 1974). Some designs allowed for the control of specimen temperature to reduce evaporation of liquid by using a Peltier cooler (Parsons et. al. 1974), which is also commonly used in ESEM (Danilatos, 1991), or by using a heat exchanger (Parsons et. al. 1974). Biological specimens are now more commonly studied by cryogenic TEM, and modern open-cell ETEMs are primarily used to study material specimens such as nanocatalysts at high spatial resolution in lower pressure gas environments (<20 Torr), rather than liquids (Helveg et. al. 2015; Bugnet et. al. 2017; Liu et. al. 2017; Luo et. al. 2017; Lawrence & Crozier, 2018; Yang et. al. 2019).

The closed-cell method is now a powerful technique for the study of specimens in liquid in TEM and has several advantages over the open cell method. Typical closed cells can withstand pressures of up to a few atmospheres (Ross, 2015). Some designs allow for heating of the liquid up to $100^{o}C$ or beyond using a built-in heater (White et. al. 2011, Denoual et. al. 2019) and can in principle allow the liquid to be cooled to cryogenic temperatures (Tai et. al. 2014). In contrast, temperature and pressure must be carefully controlled using the open cell method in order for liquid to form and remain stable (see Figure 1). However, the open-cell approach may be advantageous in some circumstances. For example, the necessary short separation between windows in the closed-cell method places geometric constraints on the specimen, which would not be necessary in an open-cell. Furthermore, the open-cell approach could potentially offer greater flexibility in terms of the position and types of detectors and external stimuli that can be used to probe and analyze the specimen. In addition, the open-cell approach eliminates the risk of a ruptured window causing a leak of liquid into high vacuum. These potential advantages have continued to motivate various approaches for open-cell in-situ liquid TEM studies. Gai designed a system in which liquid could injected onto the specimen in-situ from an external reservoir using a specially designed holder (Gai, 2002). Dai et al. have demonstrated a combined liquid encapsulation/ETEM approach that uses a water vapor atmosphere in an ETEM to prevent drying of liquid droplets located between two carbon films (Dai et. al. 2005). Ultra-low vapor pressure ionic liquids, which remain stable in a high vacuum TEM sample chamber, have been employed to perform open-cell in-situ liquid studies of lithium-ion battery materials (Huang et. al. 2010; Wang et. al. 2010), although the method is not applicable for the study of specimens in other liquid environments such as water.

Here, we demonstrate a new approach that enables differentially pumped, open-cell ETEMs to be used to conduct in-situ experiments in aqueous liquid solutions, offering an alternative to closed-cells. Our method is partly inspired by earlier research in aerosol science that demonstrated the phenomenon of deliquescence on salt particles in ETEM (Wise et. al 2005; Wise et. al. 2008). Our approach uses hygroscopic salt particles to act as nucleation sites for the formation of droplets of aqueous liquid solution upon exposure to water vapor. Specimen particles of interest may be coloaded into the TEM alongside salt particles and become enveloped by liquid droplets of solution as the salt particles are hydrated in-situ. In contrast to past open-cell liquid ETEM techniques, our use of in-situ hydration enables specimens to be loaded into the ETEM in a dry state using standard 3 mm TEM grids, and specimens can therefore be prepared using trivial sample preparation techniques. Our future goal will be to combine this open-cell in-situ liquid technique with an in-situ light source (Liu et. al. 2016) to study photocatalysis and photocorrosion in nanoparticles.

**Materials and Methods**

CdS nanoparticles with a nominal diameter of ~5 nm (US Research Nanomaterials, Inc., Houston, TX, USA) were used as a test sample. CdS nanoparticles were deposited onto a TEM sample grid alongside micron-sized ball-milled particles of NaCl (Sigma-Aldrich) using a dry preparation method.

Data was collected from the samples using an FEI Titan ETEM (Thermo-Fisher) operated at 300 kV. Up to 18 Torr of water vapour can be introduced into the ETEM sample chamber by evaporating liquid water from a glass container on a branch of the microscope's gas inlet system. The microscope was equipped with an EDAX X-ray spectrometer.

As in ESEM, the basic principle of our method involves maintaining an atmosphere of water vapour in the electron microscope sample chamber above the saturation vapour pressure at a given sample temperature, which is illustrated in Figure 1.

As was first observed in TEM by Wise et. al. (Wise et. al. 2005), particles of salt such as NaCl will deliquesce below the saturation vapor pressure of water. For NaCl, this occurs at ~76% relative humidity (Wise et. al. 2008). Our method uses this phenomenon to facilitate the formation of droplets of aqueous solution around the CdS nanoparticles. In addition, condensation of water around the salt particles below the water saturation pressure can be beneficial for imaging. This is because a higher gas pressure results in increased electron scattering from the gas, broadening the beam and reducing resolution. The vapor pressure curves in Figure 1 therefore suggest that imaging conditions will be optimized at lower temperatures, and for a given temperature, the optimum imaging conditions for imaging specimens in an aqueous solution lie in the green shaded region. In practice, the ability to achieve these conditions will be limited by the ability of the experimental apparatus to hold a constant temperature and pressure.

Figure 2a shows a schematic diagram of our experiment. In order to control sample temperature within the microscope, samples were loaded into a cryogenic side-entry TEM sample holder (Gatan). After the holder was inserted into the TEM, the sample temperature was controlled by filling the dewar of the holder, which is in thermal contact with the sample, with a frozen water/ethanol solution, which was poured into the dewar in crushed ice form. This method relies on the latent heat of the solution to hold the sample at a constant temperature whilst the solution is changing state (i.e. melting from solid to liquid). A water/ethanol solution was chosen for two main reasons. Firstly, because of the relatively large latent heat of fusion of water in particular, which leads to a longer period of time over which the temperature of the solution will be constant. Secondly, because the precise melting temperature of the solution can be optimized by varying the water/ethanol volume ratio, as shown by Figure 2b. In practice, there is a difference in temperature between the sample and the solution. This is due to the fact that the sample inside the microscope is necessarily some distance from the dewar (see Fig 2a), which is outside the microscope. The sample holder is also in thermal contact with the microscope itself, and the gas flowing into the sample chamber (i.e. water vapor), which enters the TEM at room temperature (~20$^{o}$C). The water vapor is produced by attaching a glass vial of liquid water to the gas handling system of the ETEM and allowing the water to slowly evaporate upon exposure to the low pressure. No carrier gas is required. Experimentally, we found the sample temperature (measured using a thermocouple built into the holder) to be ~ 28$^{o}$C warmer than the melting solution in the dewar (as measured using a thermometer) when gas was present (the difference was only ~18°C in vacuum). In order for the temperature of the sample to be held in the ~0-10°C range during experiments, the volume ratio of ethanol to water had to be chosen such that the melting point of the solution would be ~ between -18°C and ~ -28°C. A ratio of ~40% ethanol to 60% water was therefore chosen (see Fig 2b).

During the experiment, sample temperature was monitored using the thermocouple built into the cryo-holder. The thermocouple is located near the tip of the holder, a few mm from the TEM grid containing the specimen. If the sample temperature began to rise quickly due to the complete melting of the solution, small amounts of liquid nitrogen were added to the dewar. This would rapidly re-freeze the solution, allowing the experiments to continue once the sample temperature equilibrated. Sample temperature would typically decrease by 1-5°C, depending on the quantity of liquid nitrogen added to the dewar.

**Results and Discussion**

Figure 3a shows a cluster of CdS nanoparticles on a micron sized NaCl particle, prior to cooling the sample and the introduction of water vapour to the ETEM sample chamber. Figure 3b shows the same area of the sample, after the introduction of ~5 Torr of water vapour to the sample chamber, with a sample temperature of ~2°C. A liquid water droplet is now present, and the micron sized NaCl particle is no longer visible indicating that it has dissolved in the water droplet, forming an aqueous solution. The droplet appears to have enveloped the CdS particles that originally decorated the NaCl particle, since no CdS particles can be seen outside the exterior perimeter of the droplet. Figure 3c shows the same area again, but now with ~8 Torr of water vapor in the ETEM sample chamber, with a sample temperature of ~5 °C. The diameter of the water droplet appears to have increased, suggesting that varying the water vapor pressure in the chamber may allow a degree of control of the droplet size. In general, the size and shape of a droplet will also depend on wettability, which will be a function of both the liquid and the support material. For liquid water on an amorphous carbon support film (typical of TEM grids), wettability can vary depending how the amorphous carbon is prepared (Zhou et al 2006). Interestingly, in our experiment CdS clusters are not observed to move a significant distance within the water droplet over time. Given that all of the particle clusters are observed to be within the bounds of the water droplet, this suggests that the clusters are held in position either on the inside surface of the droplet, or by interaction with the carbon film at the base of the droplet. Further investigation will be needed to determine which mechanism is responsible for maintaining static CdS clusters.

The concentration of the aqueous solution formed as liquid dissolves the salt particle is controlled by the original mass of salt crystal and the diameter of the liquid water droplet. In-situ X-ray energy dispersive spectroscopy may be used to detect CdS particles in liquid and estimate the concentration of the NaCl solution (Figure 4). Following the acquisition of an experimental X-ray spectrum, the integrated intensity of the Cl and O peaks may be calculated. For simplicity, a linear background is subtracted before integrating each peak. This can be converted to a local mass, or atomic ratio of oxygen to chlorine, which can in turn be converted to give the number of moles of NaCl present per liter of $H_2O$, i.e. the local molarity of the aqueous solution. It is preferable to avoid using the Na peak for the calculation if the sample is loaded onto a copper TEM grid, as a Cu-L peak appears that overlaps with the Na-K peak (Figure 4b). Interpretation of the experimental spectra was aided by comparing experimental spectra to simulated spectra performed using the NIST-DTSA II program of a 6 µm hemispherical water droplet in a 5 Torr $H_2O$ atmosphere, containing varying quantities of NaCl. From the results of our simulations, we conclude that the local molarity of the NaCl solution in the area imaged in Figure 4 is ~2 M, which is ~3 times the concentration of seawater (~0.6 M) (Kozarac et. al. 1976).

Potential to improve the method

Though our method has successfully enabled in-situ formation of an aqueous solution in ETEM, there are areas for potential improvement. The main area for improvement involves control of sample temperature. We find that the frozen water/ethanol blend in the cryo-holder will keep the sample within the 1-10$^o$C temperature range necessary to maintain aqueous droplets for ~1 hour if left unattended. Experiments can continue for longer than 1 hour so long as small quantities of liquid nitrogen are added to the water/ethanol blend to cool it down and ensure that it remains frozen. However, a more stable method of sample temperature control would reduce sample drift and simplify the technique. In future, it may be more practical to use a specimen holder that has been optimized for insulating frozen water/alcohol blends, or alternatively to use a Peltier cooled specimen holder, similar to that of Parsons et. al. (Parsons et. al. 1974). Another area for improvement would be to use electron energy loss spectroscopy (EELS) to strengthen the characterization of the liquid. Imaging and X-ray energy dispersive spectroscopy are currently the only tools available for direct characterization of the liquid droplets in our ETEM. However, in future, the addition of a high energy resolution spectrometer may allow droplets to be characterized using EELS. In particular, one may be able to use EELS to detect the vibrational states of the $H_2O$ molecule, as has been demonstrated by Jokisaari et. al. (Jokisaari et. al 2018). There will of course be some contribution to the spectrum from water vapor, but one would expect the strength of the vibrational $H_2O$ signal from a given area to increase strongly when liquid is formed due to the much higher density of $H_2O$ molecules in the liquid phase.

**Conclusions and Future Work**

In summary, we have demonstrated the formation of liquid droplets of aqueous solution in an open-cell ETEM, using in-situ hydration of NaCl particles. We have shown that these droplets can envelope a material of interest that has been co-loaded with NaCl, providing an open-cell method for in-situ studies of the material in liquid. In future, we aim to use this technique in conjunction with our in-situ illumination system study the effects of photocorrosion on photocatalyst nanoparticles. Future work will also focus on further developing the technique by exploring alternative methods for controlling sample temperature, testing the limits of spatial resolution, exploring the use of aloof beam EELS to detect bandgap states within nanoparticles in the liquid droplet, and investigating the effect of different salts, and different salt particle preparation methods on droplet size. Alternatives to salt particles could be investigated to facilitate water nucleation, such as porous carbons, which would not result in solute ions in the liquid. The use of liquids other than water, such as alcohols, or battery electrolytes could potentially be explored, as could the potential of using mixed gas atmospheres to study reactions at the gas-liquid-solid interface.

Acknowledgements

We gratefully acknowledge support from the U.S. DOE (DE-SC0004954) and ASU's John M. Cowley Center for High Resolution Electron Microscopy.

Grogan, J. M. & Bau, H. H. (2010). The Nanoaquarium: A Platform for *In Situ* Transmission Electron Microscopy in Liquid Media. *Journal of Microelectromechanical Systems* **19**, 885–894.

Helveg, S., Kisielowski, C. F., Jinschek, J. R., Specht, P., Yuan, G. & Frei, H. (2015). Observing gas-catalyst dynamics at atomic resolution and single-atom sensitivity. *Micron* **68**, 176–185.

Holtz, M. E., Yu, Y., Gao, J., Abruña, H. D. & Muller, D. A. (2013). *In Situ* Electron Energy-Loss Spectroscopy in Liquids. *Microscopy and Microanalysis* **19**, 1027–1035.

Huang, J. Y., Zhong, L., Wang, C. M., Sullivan, J. P., Xu, W., Zhang, L. Q., Mao, S. X., Hudak, N. S., Liu, X. H., Subramanian, A., Fan, H., Qi, L., Kushima, A. & Li, J. (2010). In Situ Observation of the Electrochemical Lithiation of a Single $SnO_2$ Nanowire Electrode. *Science* **330**, 1515–1520.

Jensen, E., Burrows, A. & Mølhave, K. (2014). Monolithic Chip System with a Microfluidic Channel for *In Situ* Electron Microscopy of Liquids. *Microscopy and Microanalysis* **20**, 445–451.

Jokisaari, J. R., Hachtel, J. A., Hu, X., Mukherjee, A., Wang, C., Konecna, A., Lovejoy, T. C., Dellby, N., Aizpurua, J., Krivanek, O. L., Idrobo, J.-C. & Klie, R. F. (2018). Vibrational Spectroscopy of Water with High Spatial Resolution. *Advanced Materials* **30**, 1802702.

de Jonge, N. & Ross, F. M. (2011). Electron microscopy of specimens in liquid. *Nature Nanotechnology* **6**, 695–704.

Kozarac, Z., Ćosović, B. & Branica, M. (1976). Estimation of surfactant activity of polluted seawater by kalousek commutator technique. *Journal of Electroanalytical Chemistry and Interfacial Electrochemistry* **68**, 75–83.

FIGURE CAPTIONS

**Figure 1.** Saturation vapour pressure of water as a function of temperature (dark blue line). Above this pressure (blue area), liquid water may form in an electron microscope specimen chamber. The light blue line indicates the point at which deliquescence of NaCl salt particles can occur (~76% of relative humidity). In the green shaded area, liquid water can form around salt particles.

**Figure 2.** a) Schematic diagram of experimental setup, showing the cryogenic holder and the electron microscope sample chamber. b) Freezing temperature of water/ethanol solutions as a function of ethanol volume percentage (Engineering Toolbox, 2005).

**Figure 3.** a) TEM image of a typical cluster of sample particles before exposure to water vapor. b) Image of the same area after ~20 minutes exposure to > 5 Torr water vapor at ~ 2ºC, showing CdS clusters enveloped by a water droplet. The NaCl appears to have dissolved in the water, forming a solution. c) TEM image of the same area after a further 10 minutes of exposure to water vapor at ~ 8 Torr. The droplet of solution has increased in volume, but the relative positions of the CdS clusters appear unchanged. All scale bars are 2 μm.

**Figure 4.** a) TEM image of particles immersed in solution. b) X-ray spectrum from the field of view shown in (a). The presence of the sulfur K and cadmium L edges confirms that we are able to detect CdS particles spectroscopically when they are enveped in liquid. The dashed red line shows a simulated spectrum from a 6 μm hemispherical droplet of ~2 M NaCl solution, containing a block of CdS, in a 5 Torr water vapor atmosphere. The simulation matches the data reasonably well.

**Figure 1.**

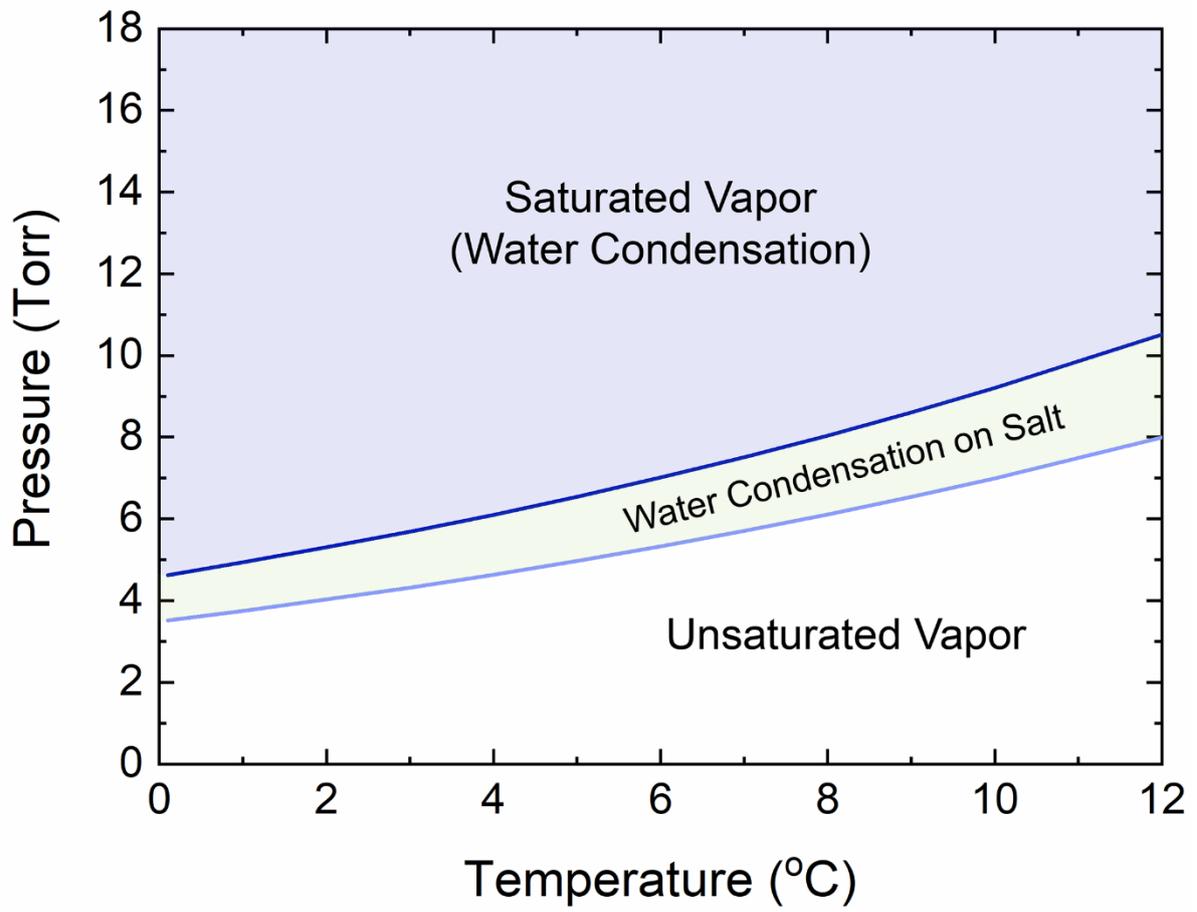

**Figure 2.**

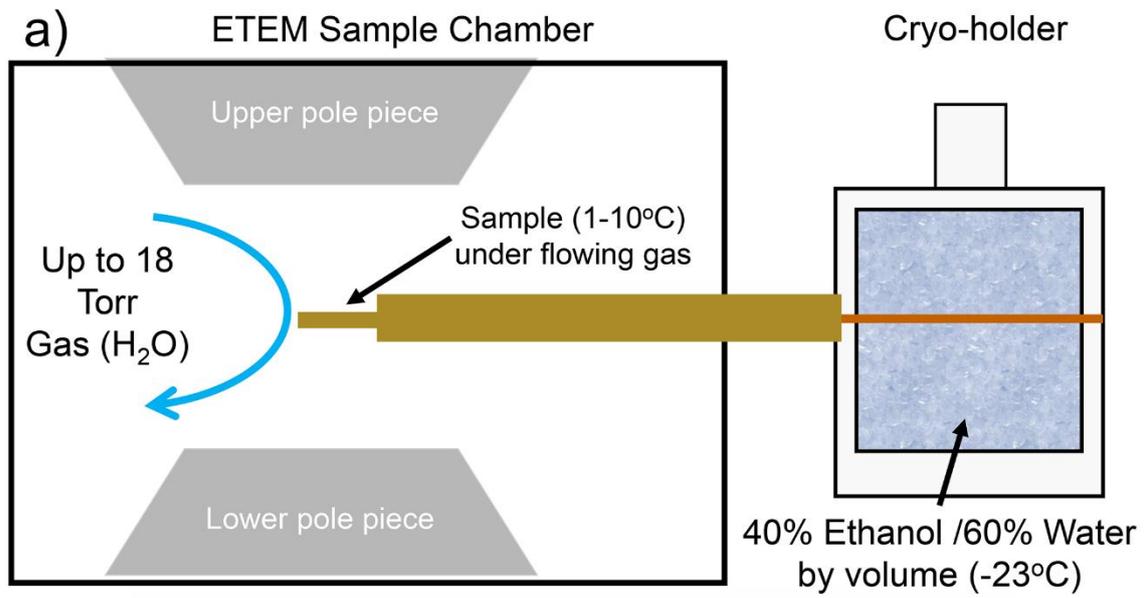

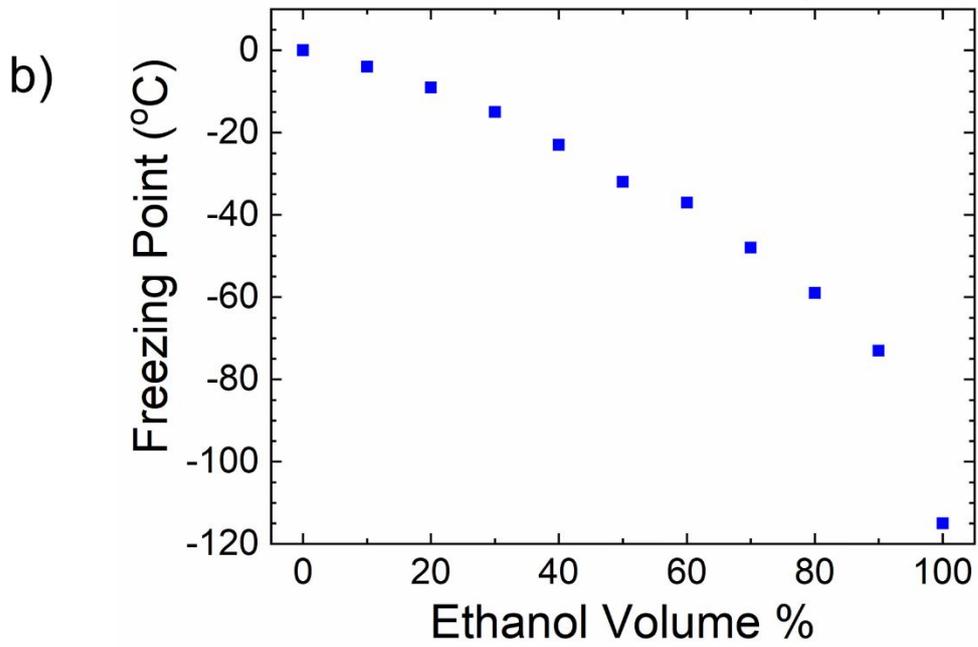

**Figure 3.**

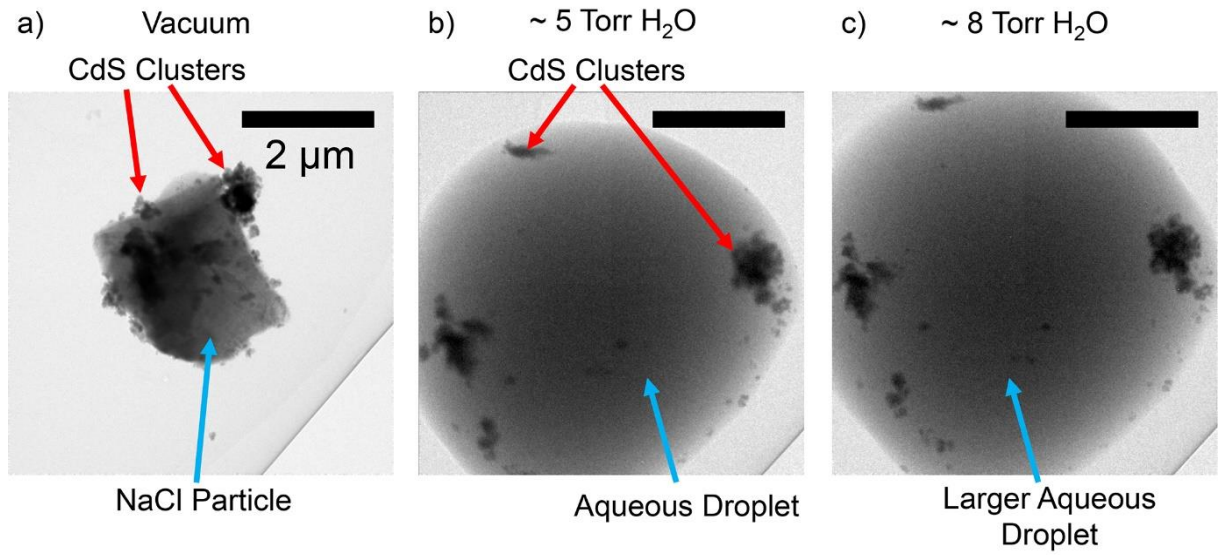

**Figure 4.**

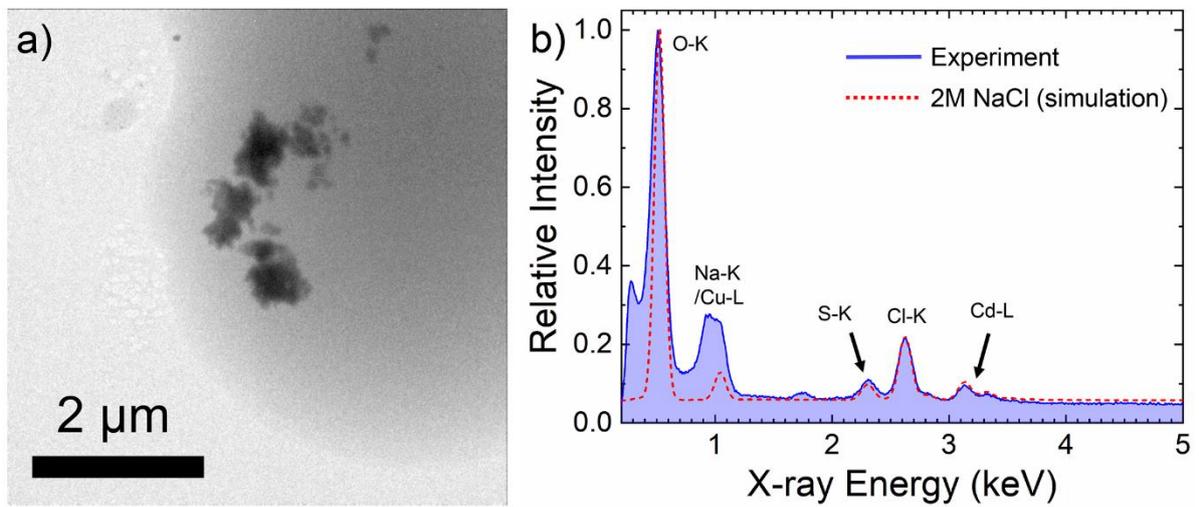